\documentclass[letterpaper,11pt]{article}

\usepackage{graphicx}
\usepackage{amsmath}
\usepackage{wasysym}
\usepackage{lscape}
\usepackage{url}

\usepackage[T1]{fontenc} 
\usepackage[applemac]{inputenc} 
\usepackage{color}
\usepackage{epstopdf}
\usepackage{setspace}  
\onehalfspace

\begin{document}
\author{A. Kellerer} 
\title{Quantum telescopes}
\date{}

\maketitle

{\bf In the $20^{\rm th\/}$ century, quantum mechanics connected the particle and wave concepts of light and thereby made mechanisms accessible that had never been imagined before. Processes such as stimulated emission and quantum entanglement have revolutionized modern technology. But even though astronomical observations rely on novel technologies, the optical layout of telescopes has fundamentally remained unchanged. While there is no doubt that Huyghens and Newton would be astounded by the size of our modern telescopes, they would nevertheless understand their optical design. 
The time may now have come to consider quantum telescopes, that make use of the fundamental scientific changes brought along by quantum mechanics. While one aim is to entertain our reader, our main purpose is to explore the possible  future evolution of telescopes.\/}

\section*{Ever larger telescopes}

\begin{figure}[htbp]
\begin{center}
\includegraphics[width=.45\textwidth]{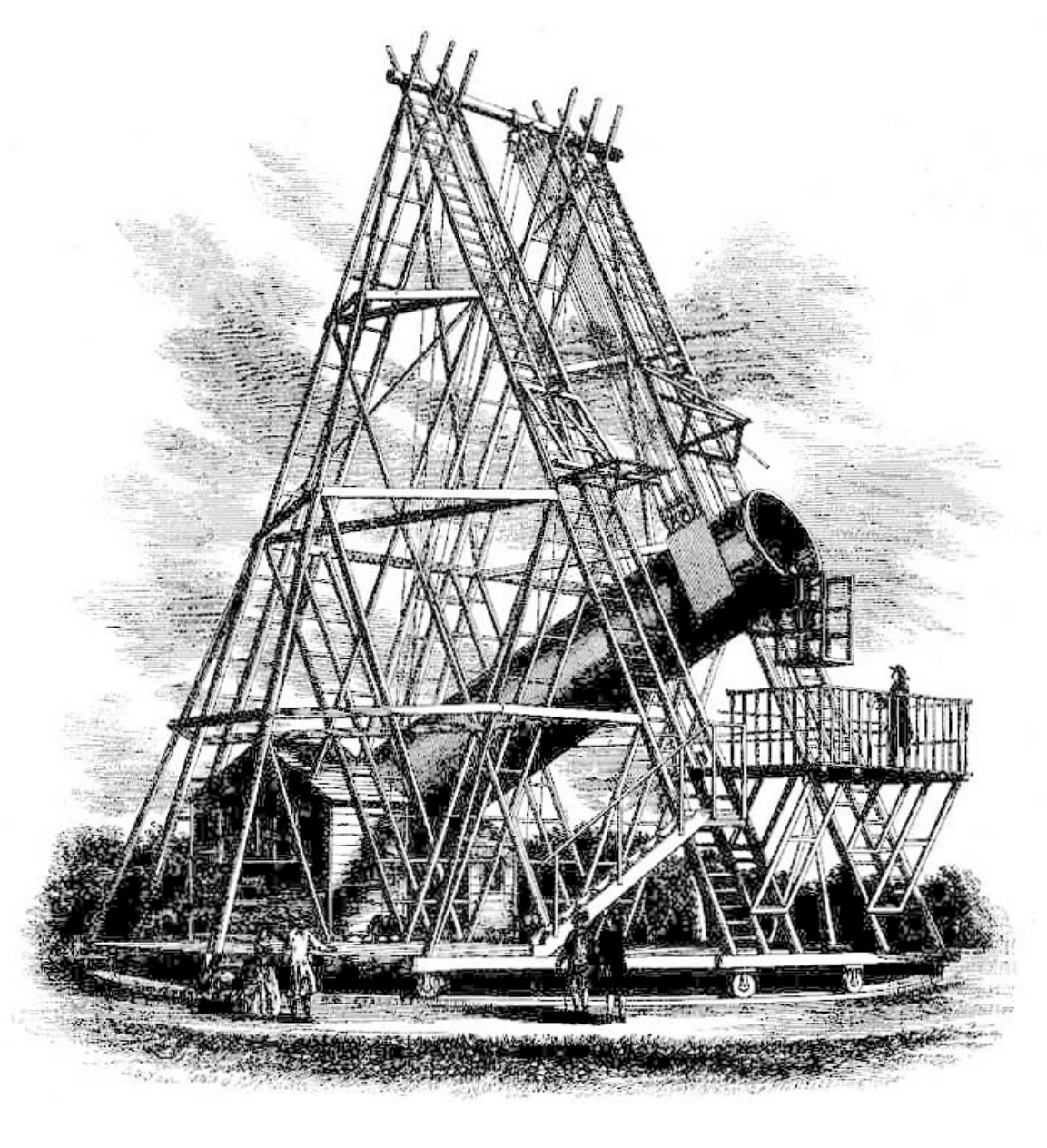}
\caption{The {\it Big Forty Foot\/} telescope in Berkshire, near Windsor castle. Built 1785-1789 by William and Caroline Herschel, it remained the world's largest telescope for over 50 years.}
\label{fig:1}
\end{center}
\end{figure}

Since the telescope was first invented, astronomers have striven to build ever larger telescopes. Image\,\ref{fig:1} shows the {\it Great Forty Foot\/}, built by William Herschel and his sister Caroline. Its aperture diameter of 1.2\,m was gigantic for the time. Accordingly, the costs greatly exceeded the initial estimates. But William Herschel had discovered a planet in 1781. He had called it the {\it Georgian Star\/} in honor of King George III, and even though the planet ultimately became to be known as Uranus, the initial appellation insured sufficient funding throughout construction. 
Eventually the telescope was completed, but it turned out to be somewhat of a disappointment: the image quality was generally poor, due to the detrimental effects of atmospheric turbulence and Herschel did most of his observations with the smaller {\it Twenty Foot\/} telescope\,\footnote{Twenty foot refers to the focal length. The aperture diameter equaled 0.5\,m.}. 
By the time, it seemed that it would always remain lost labor to build telescopes larger than about 1\,m diameter. 
Today means exist to correct for the effects of atmospheric turbulence and thus a race has begun for ever larger telescopes. In its time the Great Forty Foot remained the largest telescope for over fifty years, nowadays a new telescope breaks the record  every 4-5 years. Clearly, the contest will continue.

There are two main reasons for building larger telescopes. The first is sensitivity: larger collecting surfaces are needed to see fainter targets, much in the same way as the eye's pupil enlarges at night in order to sense fainter objects. The second reason is somewhat less intuitive: larger telescopes allow us to see smaller details on astronomical targets. The fundamental reason for this goes beyond the classical description in terms of the wave formalism, it is rooted in quantum mechanics.
Quantum mechanics describes physical phenomena at the smallest scales. Not the scale of biologic cells. Cells can be seen through a microscope, they have a position. In contrast, particles described by quantum mechanics -- e.g. electrons, protons, photons -- do not have a position. 
Their positions are extended and this is lucky, since -- if they could assume a point-like position -- the negatively charged electrons would fall into the positively charged atomic nuclei and atoms would collapse. The stability of atoms was a mystery before the advent of quantum mechanics. Nowadays the position of the electron is described by a wave function: the electron is spread out all over the wave-like structure. When a measurement is performed to localize the electron the probability to detect it around a particular point is proportional to the square of the amplitude. In the case of an electron that is bound to its atom the wave has a node in the atomic nucleus and a node outside the atom. The electron thus remains bound to the nucleus but does not fall into it. The first person who formulated the resulting uncertainty as a fundamental characteristics of quantum mechanics was Werner Heisenberg\,\cite{Heisenberg, Landau}. As we will see, this uncertainty is of direct relevance for telescopes.
 
 \section*{The strange process of astronomical imaging}
 
Imagine we point a telescope towards a distant galaxy. The galaxy emits light in form of quanta (photons). We know that light is emitted in quanta because if the target is faint enough, the detector records the arrival of the individual photons, rather than a continuous signal. Let’s consider one of these photons: It does not have a point-like position, rather it is extended over a sphere centered on the location where it was emitted in the distant galaxy. The sphere diameter increases as the photon propagates. The photon is not at one unknown point on this sphere, it is everywhere on the spherical surface. 
Counter-intuitive as it appears, this is a clear result of measurements: the photon is everywhere in this region. But once a detector records the photon, it records it entirely: it does not record half or a quarter of the photon’s energy, it records its total energy. Just before its detection the photon could have been localized anywhere else on the immense spherical surface extending several thousands, millions of light years. But when the photon is detected, the entire wavefront collapses instantly, a process that Einstein, Podolsky and Rosen in 1935 considered sufficiently absurd to prove that Bohr's and Heisenberg's interpretation of quantum mechanics was incomplete and preliminary\,\cite{EPR}. Later John Steward Bell demonstrated that any attempt to ascribe to the photon or electron a hidden classical path or location must violate the quantum mechanical predictions\,\cite{Bell}, and recent experiments, notably by Anton Zeilinger's group in Vienna\,\cite{Zeilinger1999, Zeilinger2007}, have shown that the seeming disregard of the wave function for distance in time and space is real.

To repeat: Imagine a creature on another planet around a distant star, who is simultaneously observing the same galaxy. Before the photon is detected, we both have the same chance to observe it. Once the fellow creature detects the photon, our chance to observe it vanishes. Something happens a thousand or millions of light-years away and has an immediate effect on us. 
Albert Einstein called this a spooky action at a distance\,\cite{BornEinsteinLetters}. He had determined that nothing in the Universe propagates faster than the speed of light. The mysterious collapse of the wave function disregards this limitation.

Let  us return to the telescope. Before reaching it, the photon is spread out over a large surface centered around the galaxy.  When the wave enters the telescope the uncertainty of the lateral position is reduced to the radius of the aperture, $\sigma_x=r$. In line with the Heisenberg principle there is then an uncertainty, $\sigma_x \geq \hbar/2r$,  of the lateral momentum, i.e. the initial direction cannot be retrieved with a precision better than the diffraction limit. Astronomical features that are separated by less than the diffraction limit cannot be distinguished. The solution consists in building larger telescopes. If the aperture of the telescope is increased, the knowledge of the photon position is reduced, $\sigma_x$ is enlarged and the precision of the momentum improves. The diffraction patterns become more narrow, and smaller astronomical features can be distinguished.

\begin{figure}[htbp]
\begin{center}
\includegraphics[width=.45\textwidth]{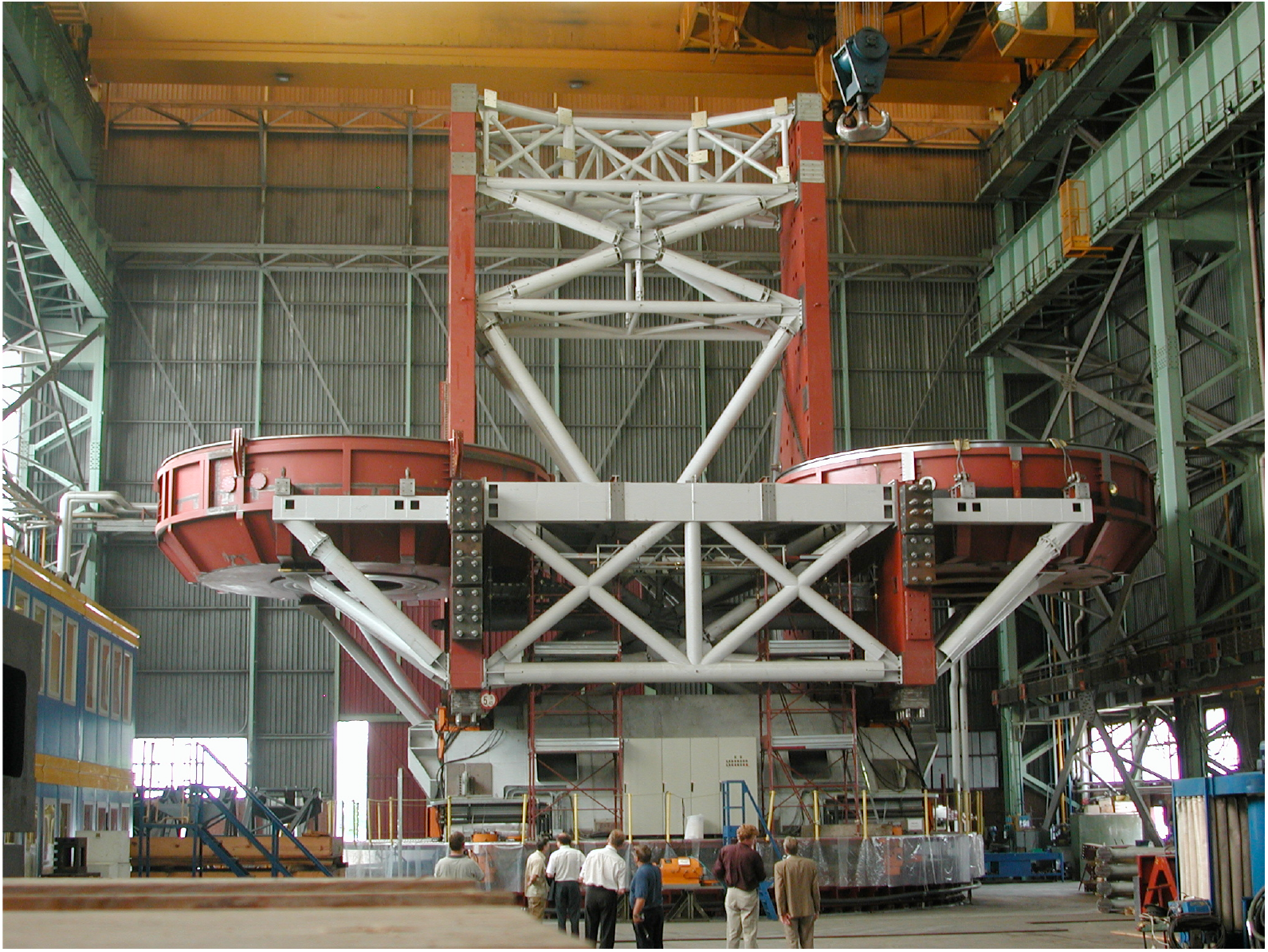}
\caption{The {\it Large Binocular Telescope\/} in Arizona, USA, is currently the largest optical telescope in the world. The light collected by the two 8.4\,m diameter mirrors is recombined on the same detector.}
\label{fig:LBT}
\end{center}
\end{figure}

\begin{figure}[htbp]
\begin{center}
\includegraphics[width=.45\textwidth]{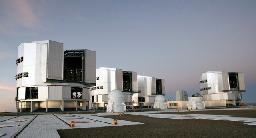}
\caption{The {\it Very Large Telescope\/} operated in Chile by the European Solar Observatory. The light collected by four 8\,m diameter mirrors (up to 130\,m apart) can be recombined on the same detector. }
\label{fig:VLT}
\end{center}
\end{figure}

Ever larger telescopes are therefore being built. Currently, the largest telescope is the {\it Large Binocular Telescope\/}: see Image\,\ref{fig:LBT}. Each of its two mirrors has an 8.4\,m diameter. An added feature is that the light collected by both mirrors is sent onto the same detector. In this way, one cannot say that the photon passed one of the two apertures: it passed both apertures. As our knowledge of the photon position is reduced, the precision of the photon momentum improves. Interferometry allows to achieve extremely high angular resolution with telescopes of moderate sizes\,\cite{Monnier, Labeyrie}. 
This is a smart solution, which has been implemented in many observatories. Another example is the {\it Very Large Telescope\/} (VLT) of the  {\it European Southern Observatory\/}: see Image\,\ref{fig:VLT}. It consists of four 8\,m diameter mirrors. The light collected by the four mirrors can be combined onto one detector. Several further smaller telescopes can likewise contribute signal. After being collected by the telescopes the light is propagated through underground tunnels and is reflected on a series of mirrors before it arrives on the final detector.  The mirror positions need to be controlled at fractions of a wavelength (i.e. with $10^{-9}$ meter precision). The interferometric recombination of light is a major technological challenge. 
In order to alleviate the problem today's telescopes are situated on dry mountain tops, where the detrimental effects of atmospheric turbulence are minimized. 

\section*{Atmospheric turbulence}
 
 \begin{figure}[htbp]
\begin{center}
\includegraphics[width=.6\textwidth]{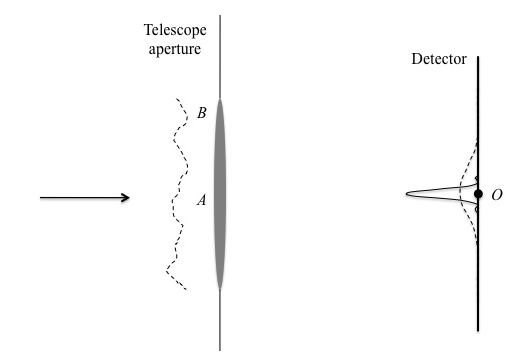}
\caption{A photon is collected by a telescope. The shape of the lens compensates exactly for the path-length differences between different points on the aperture and the focal point O on the detector. In the presence of atmospheric turbulence the wavefront is distorted and the images are blurred.}
\label{fig:flatwave}
\end{center}
\end{figure}

Image\,\ref{fig:flatwave} represents a photon collected by a telescope.   The lines in the telescope aperture-plane represent areas where the photon, i.e. the wave, is blocked from reaching the detector. The oval lens is the aperture of the telescope. Note that most telescopes use mirrors rather than lenses, because mirrors are achromatic. Here, we consider a lens to simplify the representation, since -- with a lens -- the direction of propagation is preserved.  
The distance AO is smaller than the distance BO and therefore, in the absence of the lens, the photon would take less time to go from A to O than from B to O. The index of refraction of the lens is higher than the index of refraction of air, so that the light travels slower in the lens than in the air. At point A the photon wave crosses a thick lens-section. It is thus delayed compared to point B where it passes a thinner part of the lens. This exactly compensates for the path-length difference. The shape of the lens insures that the photon takes the same time to reach point O on the detector from all points on the aperture.  
The photon has taken all virtual paths simultaneously, the position is unconstrained within the aperture and the uncertainty of the momentum is minimized. Accordingly, the angular resolution of the telescope attains its diffraction limit.

What are the effects of atmospheric turbulence? Atmospheric turbulence corresponds to variations in the optical index of air\,\cite{Roddier}.  The photon crosses some parts of the atmosphere faster than others and its wave front is therefore distorted, see dashed lines on Image\,\ref{fig:flatwave}. 
The photon is in advance at point B and delayed at point A. A precisely timed detector could in principle distinguish the events that correspond to the photon crossing the aperture at different positions. Even if we do not choose to do these measurements, the events are fundamentally distinguishable. The uncertainty on the photon position, $\sigma_x$, is reduced and thus the uncertainty on the momentum, $\sigma_p$, increases. The images recorded with the telescope are blurred, the angular resolution is degraded. 
 
  \begin{figure}[htbp]
\begin{center}
\includegraphics[width=.7\textwidth]{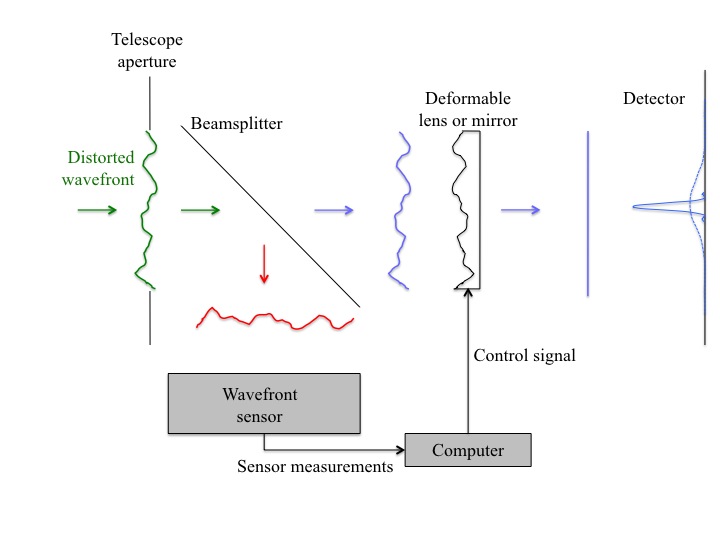}
\caption{In an adaptive-optical correction system the shape of the wavefront is sensed and a deformable mirror (represented as a lens on this diagram) uses the sensor measurements to flatten the wavefront. In advanced current systems, up to several thousand sensors control an equal number of actuators in intervals of about 1\,ms. }
\label{fig:AO}
\end{center}
\end{figure}

To mitigate the effects of atmospheric turbulence, astronomers use {\it adaptive-optical correction systems\/}\,\cite{Tokovinin}. In such a system, the distorted wavefront enters the telescope, a beam-splitter sends a fraction of the incoming photons towards a wavefront sensor, while the remaining photons are sent onto the imaging detector: see Image\,\ref{fig:AO}. 
This separation is generally done in terms of wavelength: photons of one particular wavelength are sent onto the sensor, photons of other wavelengths contribute to the high-resolution image. 
The sensor measurements are transmitted to a computer that determines voltage commands for a deformable mirror. A new correction is applied every thousandth of a second. 
Here again, we consider a deformable lens rather than a deformable mirror for ease of representation. The lens is distorted such that advanced parts of the wavefront cross thicker sections of the lens, while delayed parts cross thinner sections. Behind the lens, the wavefront emerges flat and the traversal time is thus the same from all parts of the aperture. The resolution of the telescope is then, in spite of the atmospheric turbulence, diffraction limited. 
Note that in actuality the deformable lens/mirror is placed in front of the beam-splitter, so that the correction cycle works in closed loop. We have represented an open loop for sake of simplicity.

Current projects aim at building ever larger telescopes equipped with higher-order adaptive-optical correction systems. The European Southern Observatory plans the construction of the {\it European Extremely Large Telescope\/}, a 39\,m diameter telescope to see first light in 2020 on the chilean mountain of Cerro Armazones\,\cite{EELT}. 
 
 \section*{Quantum entanglement}
 
  \begin{figure}[htbp]
\begin{center}
\includegraphics[width=.4\textwidth]{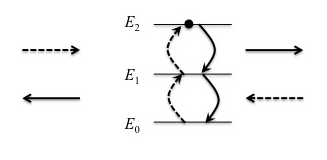}
\caption{Two photons are sent onto an atom. When they are re-emitted, they form one quantum system: they are entangled and the Heisenberg uncertainty principle applies to the ensemble of the two photons. }
\label{fig:atom}
\end{center}
\end{figure}

We have used the classical wave formalism to describe the image formation processes in telescopes. But, as quantum mechanics evolved, effects such as stimulated emission, quantum entanglement and quantum non-demolition measurements were predicted and subsequently observed (see e.g. \,\cite{Feynman, Zeilinger1999, Haroche}). Could these effects possibly be used to improve the resolution of a telescope beyond its classic diffraction limit?

Consider the following thought-experiment: Photons arrive on a crystal lattice of atoms and an atom is excited into a first and then a second excited state, as sketched on Image\,\ref{fig:atom}. From this excited state, the atom de-excites emitting two photons. The atom is part of a crystal lattice and it can therefore not appreciably recoil. Suppose that the two input photons arrived on the atom from opposite directions. The momentum being conserved the two output photons must then likewise have vanishing joint momentum, i.e. once the direction of one of the photons is measured, the direction of the other photon is constrained to the opposite direction. Upon emission both photons are spread out on the sphere that is centered on the source and that extends as the photons propagate. 
Assume one observes the atomic crystal with a telescope. When one photon is detected its wavefront collapses. What about the other photon? It has not been detected, and one might therefore expect it to still be spread out on its spherical wavefront. But this is not so: it is now constrained to the opposite of the measured direction; the detection of one photon co-determines the position of the other photon. Even if it has not been detected, the direction of the second photon is defined. 
The distance between the crystal and the telescope might be several thousand light years, nevertheless the detection of one photon collapses the wave function of the second photon onto a directionally defined ray in the Universe. The system of two photons acts as a whole, as if it had not been separated in space. This specific thought-experiment let Albert Einstein coin the expression of spooky action at a distance\,\cite{BornEinsteinLetters}.
 
Quantum entanglement is actually used to beat the diffraction limit in optical lithography\,\cite{Boto}. The aim is to engrave ever smaller details on printed circuit boards and thus to decrease the size of electronic devices. From a classic optics point of view the smallest details are set by the diffraction limit. This limit however can be overcome when photons are entangled: Before the photons get entangled via an atomic interaction, the Heisenberg uncertainty principle applies to the two photons separately. Once the photons are entangled, the uncertainty principle applies to the ensemble of the two photons. This entangled system has twice the energy of the individual photons and the diffraction limit of their mean position is overcome by a factor 2. If $N$ photons are entangled, the diffraction limit is overcome by a factor $N$\,\cite{Mitchell, Sewell}. Can this or a similar process be used to improve the resolution of the telescope?

\section*{Photon cloning}
  
We have seen that a photon collected from a distant astronomical source is diffracted upon passage through the telescope aperture. 
The photons collected from a point-like source are, therefore, distributed over an {\it Airy pattern\/} of angular width $\lambda/D$, where $\lambda$ is the spectral wavelength of the photons and $D$ is the telescope diameter. 
This width limits the angular resolution of the telescope and the astronomers dream of infinitely thin Airy patterns. 
The most straightforward means to improve the angular resolution is to build larger telescopes and thus to increase the value $D$. 
Could one, however, envisage a way to increase the angular resolution without increasing the size of the telescope\,\cite{Kellerer}?

\begin{figure}[htbp]
\begin{center}
\includegraphics[width=.8\textwidth]{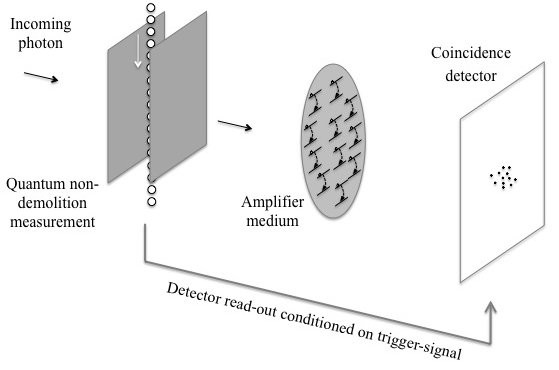}
\caption{An active telescope. Every incoming photon stimulates the emission of clones by the excited atoms. The detrimental effect of spontaneous emissions is minimized by a trigger signal, implemented via a quantum non-demolition measurement.}
\label{fig:setup}
\end{center}
\end{figure}

Imagine that we place excited atoms on a pupil plane of the telescope\,\footnote{A pupil plane is an optical image of the telescope aperture.}. The excited atoms de-excite and in the presence of a photon they tend to emit a clone of it, i.e. a photon in the same quantum state. 
With less probability they emit a photon in another quantum state, a photon emitted anywhere within $4\pi$ steradian.  
Let $N_C$ be the number, after passage through the pupil, of clones and let $N_S$ be the number of stray photons, i.e. photons in random direction. 
A coincidence detector registers the positions of simultaneously arriving photons, and the average of these values  is used as signal. If the averaging extends over the entire field, it includes the positions of the $N_C$ clones as well as the $N_S$ stray photons. It will then follow a broad distribution of width not better then $\lambda/D$, as in the absence of amplification. The Heisenberg uncertainty relation is thus preserved upon optical amplification\,\cite{Caves}: the minimum amount of noise, due to the stray photons  is such that the gain in resolution due to stimulated emissions is just offset; as Richard Feynman puts it: ``Natures has got it cooked up so we'll never be able to figure out how She does it''\,\cite{QED}.

However, as one integrates over a reduced field, less stray photons contribute to the signal. In the limit, where only clones contribute to the signal the distribution would be more narrow by a factor $\sqrt{N_C}$ than the classical diffraction width $\lambda/D$. Even though the Heisenberg uncertainty relation is not overcome in total, it is overcome on the reduced fraction of clones, $N_C$. 

The remaining draw-back of such a setup is that excited atoms emit spontaneously also in the absence of a photon from the astronomical target. These spontaneous emissions contribute an additional, extraneous  noise factor which becomes smaller as the viewing angle decreases, but may still be prohibitive under practical circumstances. To reduce this added noise, a trigger signal might be used: the photon is sent through a quantum non-demolition measurement stage, in line with details given below. Such a stage senses the arrival of a photon without destroying it. The photon then passes the cavity where it stimulates the emission of clones as described above. See also Fig.\,\ref{fig:setup}. 
Only those sets of photons that are timed together with the incoming photon are utilized for imaging, the sets of photons initiated by spontaneous emission are disregarded.

\begin{figure}[htbp]
\begin{center}
\includegraphics[width=.32\textwidth]{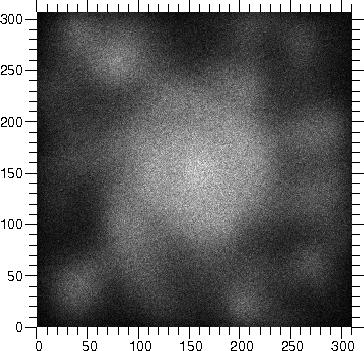}
\includegraphics[width=.32\textwidth]{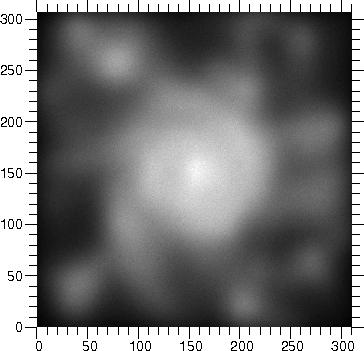}
\includegraphics[width=.32\textwidth]{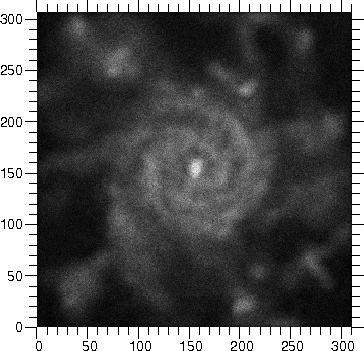}
\caption{{\bf Left panel:\/} Image of a spiral galaxy through a classical telescope. The width of the diffraction pattern equals $\lambda/D=30$\,pixels. {\bf Middle panel:\/} Each photon produces a set of 35 clones. In the absence of a coincidence detection the angular resolution is not improved. {\bf Right panel:\/} The average position of the 36 photon clones is kept as the signal -- the angular resolution is improved by a factor 6. In this numerical simulation the contribution from spontaneous emissions is neglected. The initial high-resolution image was obtained from\,\cite{Schaye}.}
\label{fig:image}
\end{center}
\end{figure}

To repeat the essential point. In any registered set of simultaneously arriving photons,  the detector location of each photon has a standard deviation according to the classical resolution. Since the coincidence mode of the detector permits the identification of the photons that belong to the same set their average -- over a chosen angular region --  can be utilized as signal, and this signal has smaller standard deviation up to the factor $\sqrt{N_C}$. Without the coincidence technique one could not obtain the mean location of the photon clones and the classic limit of the resolution would not be overcome: see Fig.\,\ref{fig:image}.

\paragraph{Quantum non-demolition (QND) measurements:\/} When photons arrive on a detector -- e.g. the retina or a CCD chip -- they interact with atoms, the energy of the photon is transmitted to the atom and the photon is thereby destroyed. The same applies to the usual discrete photon counting procedures;  as Serge Haroche puts it: ``The photon, like the Marathon soldier, was dying delivering its message''\,\cite{Cartwright}. 

In a QND measurement, the photon is detected, but not destroyed\,\cite{Braginsky, Nogues}. This uses quantum entanglement: The photon interacts with a quantum probe, typically an atom in a highly excited state. The presence of the photon changes the polarization state of the atom, but it does not de-excite the atom; the Marathon soldier is counted but survives unharmed to deliver his message. In other words, the photon keeps its momentum. 
The polarization state of the atom is measured and the outcome of the measurement determines the presence or absence of a photon. Since the probe and the photon are entangled, the state of the photon is likewise constrained, but, crucially, the photon is not destroyed. 
If the constraint  is weaker than the constraint introduced by the passage through the telescope aperture, the photon's state is not even modified. 
This then serves as an ideal trigger measurement. The photon continues its way through the telescope: it passes the medium of excited atoms, stimulates the emission of clones and the set of identical photons arrives on the detector.
 The detector read-out is conditioned on the trigger signal.  Note that the signal registered by the QND measurement need not arrive at the detector faster than the photon: the detector is read out continuously and the signals of interest are sorted out by subsequent processing.  
 
 If $N_C$ photon clones arrive on the detector simultaneously the classic diffraction limit is overcome by the factor $\sqrt{N_C}$, thus the resolution could in principle be increased arbitrarily. The price to pay, however, is loss of efficiency: the larger the desired enhancement factor $\sqrt{N_C}$, the smaller the fraction of incoming photons that produce a sufficient number $N_C$ of clones. Thus the angular resolution is improved at the price of larger exposure times, which might still be a worthwhile bargain in earth-bound or space-based astronomy.

 \section*{Outlook}
 
The set-up that has been outlined may turn out to be impractical. 
But other methods can certainly be imagined to overcome the diffraction limit on a telescope. It is therefore not our aim to suggest this particular setup. The issue of interest are the intriguing properties of light as revealed by quantum mechanics and their conceivable implications for astronomy. Elementary particles do not experience space and time as we do on our scales. Does space and time even exist for elementary particles? Or do space and time in our familiar conception merely emerge in a larger network of particles?

Quantum mechanics predicts effects that had never been imagined with the classical wave-formalism of light, effects that can be extremely counter-intuitive. As Michio Kaku says: ``It is often stated that of all the theories proposed in this century, the silliest is quantum theory. In fact, some say that the only thing that quantum theory has going for it is that it is unquestionably correct.''\,\cite{Kaku}. Today’s telescopes still rely solely on classic processes, such as the diffraction and interference of light, that are well explained by the wave-formalism. But this will change and intensity interferometry, developed by Hanburry-Brown and Twiss, can already be mentioned as an example: it relies on quantum mechanics to explain correlations in light intensities\,\cite{BT1, BT2, BT3, BT4}. However, while intensity interferometry uses the quantum mechanical characteristics of light, it does not improve upon the diffraction limit. Since intensity interferometry was proposed in 1957 further advances in quantum optics have made it possible to overcome the diffraction limit in microscopes\,\cite{STED} and also in lithography\,\cite{Boto, Mitchell}. Eventually processes such as stimulated emission, quantum entanglement and quantum non-demolition measurements  may allow to overcome the classic diffraction limit in astronomy and thus to obtain high-angular resolution even with small single-dish telescopes. If the future of data processing lies in quantum computers, the future of astronomical imaging lies in quantum telescopes.

\bibliographystyle{abbrv}
\bibliography{References}
 
\end{document}